\documentclass{PoS} 

\usepackage{boxedminipage}
\usepackage{amssymb}
\usepackage{fontenc}
\usepackage{times}
\usepackage{mathptmx}
\usepackage{graphicx}
\usepackage{wrapfig}

\title{PLQCD library for Lattice QCD on multi-core machines} 

\ShortTitle{PLQCD} 

\author{\speaker{A. Abdel-Rehim}, $^a$ C. Alexandrou, $^{a,b}$ N. Anastopoulos,$^c$ G. Koutsou,$^a$  I. Liabotis $^d$ and N. Papadopoulou$^c$\\ 
        \llap{$^a$}The Cyprus Institute, CaSToRC, 20 Konstantinou Kavafi Street, 2121 Aglantzia, Nicosia, Cyprus\\
        \llap{$^b$}Department of Physics, University of Cyprus, P.O. Box 20537, 1678 Nicosia, Cyprus\\
        \llap{$^c$}Computing Systems Laboratory, School of Electrical and Computer Engineering,
                   National Technical University of Athens, Zografou Campus, 15773 Zografou, Athens, Greece \\
        \llap{$^d$}Greek Research and Technology Network, 56 Mesogion Av., 11527, Athens, Greece \\            
        E-mail: \email{a.abdel-rehim@cyi.ac.cy}, \email{c.alexandrou@cyi.ac.cy}, \email{g.koutsou@cyi.ac.cy},
                \email{anastop@cslab.ece.ntua.gr}, \email{iliaboti@grnet.gr}, \email{nikela@cslab.ece.ntua.gr}} 

\abstract{PLQCD is a stand--alone software library developed under
  PRACE for lattice QCD. It provides an implementation of the Dirac
  operator for Wilson type fermions and few efficient linear
  solvers. The library is optimized for multi-core machines using a
  hybrid parallelization with OpenMP+MPI. The main objectives of the
  library is to provide a scalable implementation of the Dirac
  operator for efficient computation of the quark propagator. In this contribution, a
  description of the PLQCD library is given together with some
  benchmark results.}

\FullConference{31st International Symposium on Lattice Field Theory LATTICE 2013\\ 
                July 29 $-–$ August 3, 2013\\ 
                Mainz, Germany} 

\begin{document} 

\section{Introduction} 
Computer hardware for commodity clusters as well as supercomputers has
evolved tremendously in the last few years. Nowadays a typical compute 
node has between 16 and 64 cores and possibly an
accelerator such as a Graphics Processing Unit (GPU) or lately an
Intel Many Integrated Core (MIC) card. This trend of packing many
low-powered but massively parallel processing units is expected to
continue as supercomputing technology pursues the Exascale
regime. The current technology trends indicate that bandwidth to main
memory will continue to lag behind computational power, which requires
a rethinking of the design of lattice QCD codes such that they can
efficiently run on such architectures. Taking this into account,
PRACE \cite{prace-2ip-wp8} allocated resources for community code
scaling activities in many computationally intensive areas including
lattice QCD. The work presented here was developed under PRACE
focusing on scaling codes for multi-core machines.
The work we present deals with community codes, and more specifically
on certain computationally intensive kernels in these codes, in order
to improve their scaling and performance for multi-core
architectures. We have carried out optimization work on the 
tmLQCD \cite{tmLQCD-ref,tmLQCD-github} code and have developed a new hybrid
MPI/OpenMP library (PLQCD) with optimized implementations of the
Wilson Dirac kernel and a selected set of linear solvers. 
Our partners in this project have
also performed optimization work for the
Molecular Dynamics integrators used in Hybrid Monte Carlo codes, and also for Landau gauge fixing.
This was done within the Chroma software suite \cite{Chroma-software} and will not be discussed 
here (See \cite{D8.3} for more information). Many other community codes of course exist but
were not considered in this work (See \cite{Albert-plenary} for an
overview). 

In what follows, we will first present the work carried out for the
case of PLQCD, where we implemented the Wilson Dirac operator and
associated linear algebra functions using MPI+OpenMP. In addition to
using this hybrid approach for parallelism, we also implement
additional optimizations such as overlapping communication and
computation, using compiler intrinsics for vectorization as well as
implementing the new Advanced Vector Instructions
\cite{Intel-developer-manual}(AVX for Intel or QPX for
Blue/Gene Q) that became recently available in new generation of
processors such as the Intel Sandy-Bridge. The work done for the case
of the tmLQCD package will then be presented, where we implemented
some new efficient linear solvers, in particular those based on
deflation such as the EigCG solver \cite{eigCG}, for which we will
give some benchmark results.

\section{Dirac operator optimizations}
A key component of the lattice Dirac operator is the hopping part given by Eq.\ref{hopping-op}.
\begin{equation}
 \psi(x) = \sum_{\mu=0}^3 [ U_\mu(x) (1-\gamma_\mu) \phi(x+\hat{e}_\mu) + U_\mu^\dagger(x-\hat{e}_\mu) (1+\gamma_\mu) \phi(x-\hat{e}_\mu) ],
 \label{hopping-op}
\end{equation}
where, $U_\mu(x)$ is the gauge link matrix in the $\mu$ direction at site $x$, $\gamma_\mu$ are the Dirac matrices and $\hat{e}_\mu$ is a unit vector in the
$\mu$ direction. $\phi$ and $\psi$ are the input and output spinors respectively. Equation \ref{hopping-op} can be re-written in terms of two auxiliary fields $\theta^+_\mu(x)=(1-\gamma_\mu) \phi(x)$ and  
$\theta_\mu^-(x)=U_\mu^\dagger(x) (1+\gamma_\mu) \phi(x)$ as
\begin{equation}
 \psi(x) = \sum_{\mu=0}^3 [ U_\mu(x) \theta_\mu^+(x+\hat{e}_\mu) + \theta_\mu^-(x-\hat{e}_\mu) ].
 \label{hopping-op-half-spinor}
\end{equation}
Because of the structure of the $\gamma$ matrices, only the upper two spin components of $\theta_\mu^\pm$ need to be computed because
the lower two spin components are related to the upper ones \cite{DDHMC}. In the following we describe some of the optimizations performed for the
hopping matrix.

\subsection{Hybrid parallelization with MPI and OpenMP}
\begin{figure}[htbp]
 \begin{center}
  \includegraphics[height=0.36\textwidth]{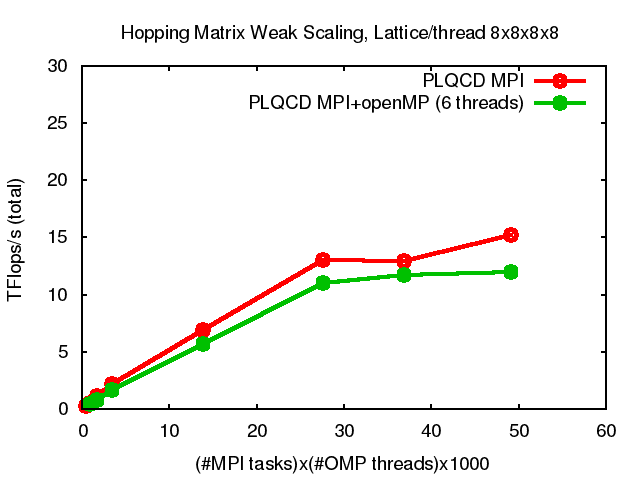}
  \includegraphics[height=0.36\textwidth]{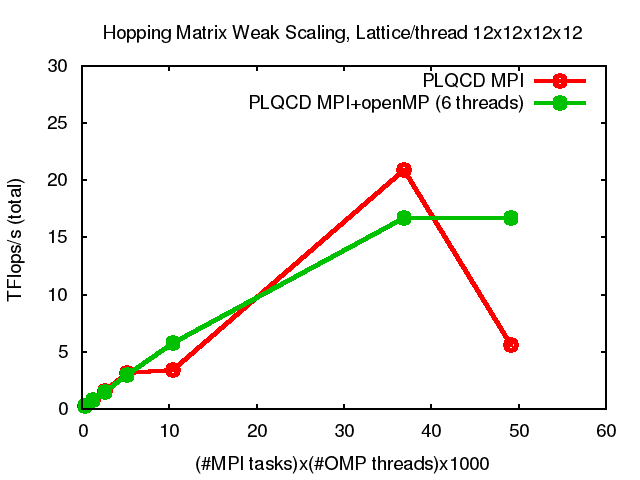}
 \end{center}
 \caption{Weak scaling test for the hopping matrix on a Cray XE6 machine with local lattice volume per core
          $8^4$(left) and $12^4$(right).}
 \label{compare-mpi-openmp}
\end{figure}

OpenMP provides a simple approach for multi-threading since it is
implemented as compiler directives. One can incrementally add
multi-threading to the code and also use the same code with
multi-threading turned on and off. Since the main component in the
hopping matrix (Dirac operator) is a large ``for loop'' over lattice
sites, it is natural to use the for-loop parallel construct of openMP.
The performance of the hybrid code is then
tested against the pure MPI version. We perform a {\it weak scaling}
test by fixing the local volume per core (or thread) and increase the
number of MPI processes. The test was done on the Hopper machine at
NERSC which is a Cray XE6\cite{hopper}. 
Each compute node has 2
twelve-core AMD 'MagnyCours' at 2.1-GHz such that each 6 cores share
the same cache. 
We find performance for the Hybrid
version is maximum when assigning at most 6 threads per MPI process
such that these 6 OMP threads share the same L3 cache. In Fig.
\ref{compare-mpi-openmp} we show the performance of the pure MPI and
the MPI+openMP with 6 threads per MPI process for a total number of
cores up to 49,152 cores. From these results we first notice that
using OpenMP leads to a slight degradation in performance as compared
to the pure MPI case. However, as we see in the case with local volume
of $12^4$, the hybrid approach performs better as we go to a large
number of cores. Similar behavior has been also observed for other
codes from different computational sciences (see the case studies on
Hopper \cite{hopper-docs-combine-omp-mpi}).

\subsection{Overlapping communication with computation}
\begin{figure}[htbp]
\begin{center}
 \includegraphics[width=0.329\linewidth]{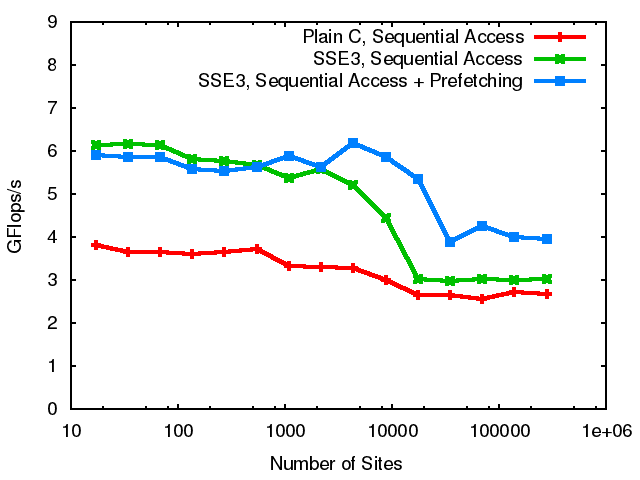}
 \includegraphics[width=0.329\linewidth]{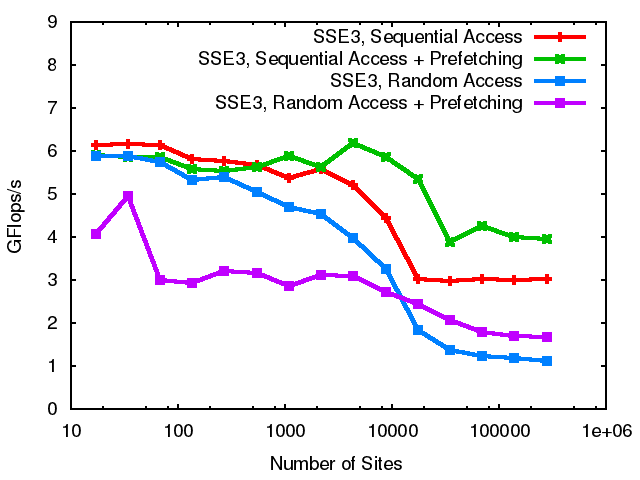}
 \includegraphics[width=0.329\linewidth]{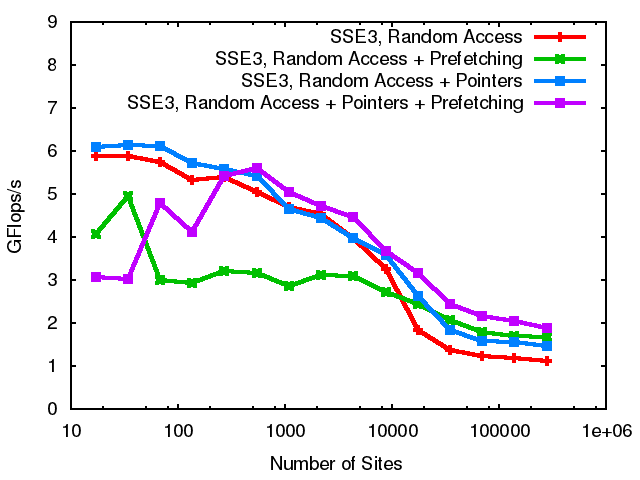}
\end{center}
\caption{Effect of prefetching for the cases of sequential (left) and
  random access (centre) of lattice sites. In the right panel we
  compare the effect of having precalculated pointer arrays for the
  case of random access.}
\label{effect-of-prefetching}
\end{figure}

Typically in lattice codes one first computes the auxiliary
half-spinor fields $\theta_\mu^\pm$ as given in Equation
(\ref{hopping-op-half-spinor}) and then communicates their values on the
boundaries between neighboring processes in the $+\mu$ and $-\mu$
directions. In a blocking communication scheme, computation halts
until communication of the boundaries completes. An alternative
approach is to overlap communications with computations by dividing
the lattice sites into bulk sites, for which nearest neighbors are
available locally, and boundary sites, for which the nearest neighbors
are located on neighboring processes, and therefore can only be
operated upon after communication. The order of operations for
computing the result $\psi$ is then done as follows:
\begin{itemize}
 \item Compute $\theta_\mu^+$ and begin communicating them to the neighboring
   MPI process in the $-\mu$ direction.
 \item Compute $\theta_\mu^-$ and begin communicating them to the neighboring
   MPI process in the $+\mu$ direction.
 \item Compute the result $\psi(x)$ on the bulk sites 
   while the neighbors are being communicated.
 \item Wait for the communications in the $-\mu$ directions to finish,
   then compute the contributions $\sum_{\mu=0}^3 [ U_\mu(x)
     \theta^+_\mu(x+\hat{e}_\mu)]$ to the result on the boundary
   sites.
 \item Wait for the communications in the $+\mu$ directions to finish,
   then compute the contributions $\sum_{\mu=0}^3 [
     \theta^-_\mu(x-\hat{e}_\mu)]$ on the boundary sites.
\end{itemize}
Communication is done using non-blocking MPI functions {\tt
  MPI\_Isend, MPI\_Irecv} and {\tt MPI\_Wait}.  
A possible drawback of this
approach is that one will access $\psi(x)$ and $U_\mu(x)$ in an
unordered fashion different from the order it is stored in
memory. This, however, can be circumvented partially by using hints in
the code for prefetching. We have tested the effect of prefetching in
case of sequential and random access of spinor and link fields. The
test was done using a separate benchmark kernel code which isolates
the link-spinor multiplication. As can be seen in
Fig. \ref{effect-of-prefetching}, prefetching becomes important for a
large number of sites, i.e.  when data (spinors and links) can not fit
in the cache memory, which is a typical situation for lattice
calculations. It is also noted that accessing the sites randomly
reduces the performance, as would be expected. In this case one can
improve the situation by defining a pointer array, e.g. for the
spinors $\tt{\psi'(i)=\&\psi(x[i])}$ where $\mathtt{x[i]}$ is the site
to be accessed at step $\tt i$ in the loop such as we show in
pseudo-code in Fig.~\ref{using-pointers-in-random-access}. These
pointers can be defined \textit{a priori}. This improves the
predictive ability of the hardware as is shown in
Fig.~\ref{effect-of-prefetching} where we compare the different
prefetching and addressing schemes.
\begin{figure}[htbp]
\begin{center}
\scalebox{0.9}{\fbox{
\begin{minipage}[t]{0.2\textwidth}
\begin{tabbing}
x\=xxx\=xxx\=xxx=\kill
Sequential access\\
\> \tt for(i=0; i< N; i++) \\
\>  \> $\mathtt{\psi(i) = U(i)\phi(i)}$
\end{tabbing}
\end{minipage}
\begin{minipage}[t]{0.2\textwidth}
\begin{tabbing}
x\=xxx\=xxx\=xxx=\kill
Random access\\
\> \tt for(i=0; i< N; i++) \\
\>  \> $\mathtt{\psi(x[i]) = U(x[i])\phi(x[i])}$
\end{tabbing}
\end{minipage}
\begin{minipage}[t]{0.2\textwidth}
\begin{tabbing}
x\=xxx\=xxx\=xxx=\kill
Random access with pointers\\
\> \tt /* prepare pointer arrays */\\
\> \tt for(i=0; i< N; i++) \\
\> \> $\mathtt{\psi'(i)=\&\psi(x[i])}$ \\
\> \> $\mathtt{U'(i)=\&U(x[i])}$ \\
\> \> $\mathtt{\phi'(i)=\&\phi(x[i])}$ \\
\> \tt for(i=0; i< N; i++) \\
\>  \> $\mathtt{*\psi'(i) = *U'(i)*\phi'(i)}$
\end{tabbing}
\end{minipage}
}}
\caption{Pseudo-code describing the site access for the case of simple
  sequential access (left), random access using an index array
  $\mathtt{x[i]}$ (centre) and random access using precalculated pointer arrays (right).}
\label{using-pointers-in-random-access}
\end{center}
\end{figure}


\subsection{Vectorization using AVX instructions}
\begin{figure}
  \begin{minipage}{0.48\linewidth}
    \begin{center}
      \includegraphics[width=1\linewidth]{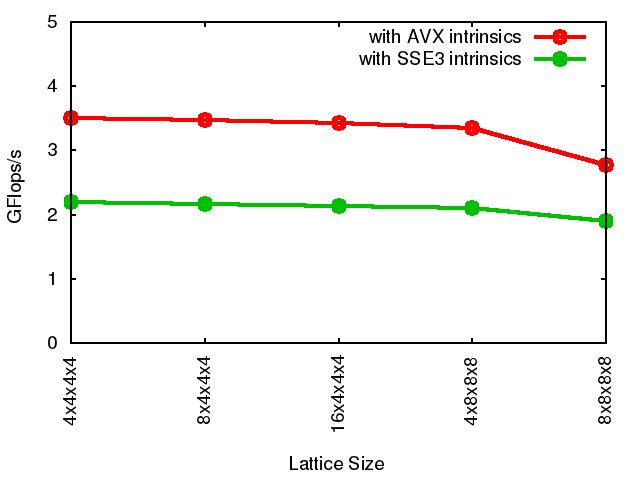}
    \end{center}
    \caption{Comparing the performance of the hopping matrix of tmLQCD
      using SSE3 and AVX in double precision on an Intel SandyBridge
      processor.} 
    \label{avx-gain}
  \end{minipage}
  \hfill
  \begin{minipage}{0.48\linewidth}
    \begin{center}
      \scalebox{0.7}{
      \fbox{
        \begin{minipage}[t]{\textwidth}
          \begin{tabbing}
            x\=xxx\=xxx\=xxx=\kill
            \tt
            \tt\#include <immintrin.h>\\
            \color{blue}
            \tt /* t0: a+b*I, e+f*I  and   t1: c+d*I, g+h*I \\
            \color{blue} \tt  * return: (ac-bd) + (ad+bc)*I,\\
            \color{blue} \tt  * (eg-fh) + (eh+fg)*I */ \\ 
            \tt static inline \_\_m256d \\
            \tt complex\_mul\_regs\_256(\_\_m256d t0, \_\_m256d t1)\\
            \{ \\
            \> \tt    \_\_m256d t2; \\
            \> \tt    t2 = t1; \\
            \> \tt    t1 = \_mm256\_unpacklo\_pd(t1,t1); \\
            \> \tt    t2 = \_mm256\_unpackhi\_pd(t2,t2); \\
            \> \tt    t1 = \_mm256\_mul\_pd(t1, t0); \\
            \> \tt    t2 = \_mm256\_mul\_pd(t2, t0); \\
            \> \tt    t2 = \_mm256\_shuffle\_pd(t2, t2, 5); \\
            \> \tt    t1 = \_mm256\_addsub\_pd(t1, t2); \\
            \> \tt   return t1;\\
            \tt\} 
          \end{tabbing}
        \end{minipage}
      }}
      \caption{Multiplying two complex numbers by two complex number of type double using AVX instructions.}
      \label{avx-code-snippet}
    \end{center}
  \end{minipage}
\end{figure}

Lattice QCD codes benefit from the vectorization capabilities of
modern processors starting from SSE to SSE3. In SSE3, we have 16
vector registers known as XMM registers, each 128-bit wide thus able
to store either 2 double precision floating point variables or 4
single precision. The AVX extensions are extensions to the x86
instruction set implemented both by Intel and AMD. These first
appeared in 2011 in the Intel SandyBridge processors and later by AMD
in their Bulldozer processor. The 16 XMM registers of SSE3 are now
256-bit wide and known as YMM registers. AVX-capable floating point
units are able to perform on 4 double precision floating point numbers
or 8 single precision. Implementing these extensions in the vectorized
parts of lattice codes has the potential of providing a gain of up to
a factor 2 in an ideal situation, although in practice this depends on
the layout of lattice data. 
We provided an implementation of
these extensions using inline intrinsics. In this implementation a
single SU(3) matrix multiplies two SU(3) vectors simultaneously. A
gain of about a factor of 1.5 is achieved for the hopping matrix in the
tmLQCD code in double precision as shown in Fig.(\ref{avx-gain}). 
For illustration, a code snippet for multiplying two complex
numbers by two complex numbers using AVX is shown in Fig.(\ref{avx-code-snippet}).

\begin{figure}[htbp]
\end{figure}
\section{EigCG solver for Twisted-Mass fermions}
\begin{figure}
\begin{center}
  \includegraphics[height=0.25\textheight]{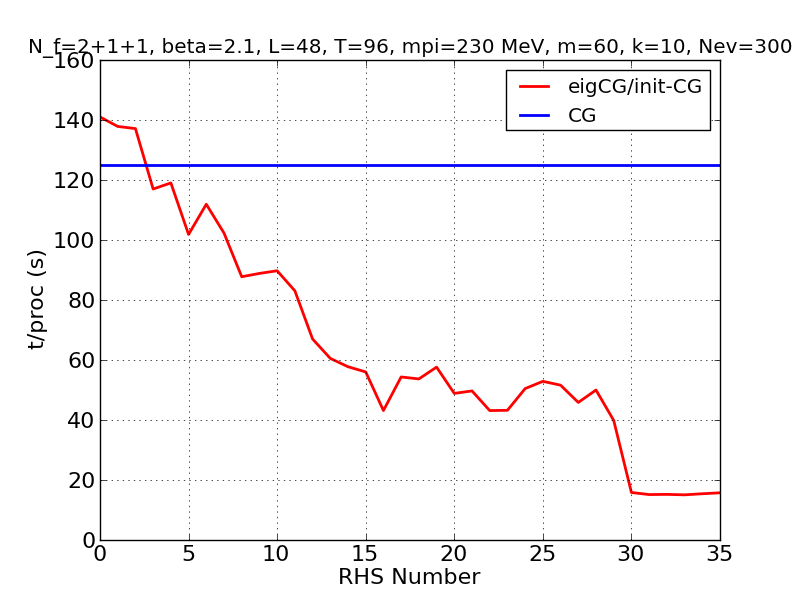}
 \end{center}
 \caption{Solution time per process for the first 35 right-hand sides
   using Incremental EigCG as compared to CG on a Twisted-Mass
   configuration with lattice size $48^3\times96$ at $\beta=2.1$, and
   pion mass $\approx 230$ MeV.}
 \label{eigcg-timings}
\end{figure}

Twisted-Mass fermions offer the advantage of automatic $O(a)$
improvement when tuned to maximal twist \cite{TMF}. Within this
development work we have added an incremental deflation algorithm,
known as EigCG, to the tmLQCD package. Numerical tests showed a
considerable speed-up of the solution of the linear systems on the
largest volumes simulated by the European Twisted Mass
Collaboration (ETMC). For illustration, we show in Fig.(\ref{eigcg-timings})
the time to solution with EigCG on a Twisted-Mass configuration with
2+1+1 dynamical flavors with lattice size $48^3\times96$ at
$\beta=2.1$, and pion mass $\approx 230$~MeV. In this case the total
number of eigenvectors deflated was 300 which was built incrementally
by computing 10 eigenvectors during the solution of the first 30
right-hand sides using a search subspace of size 60. All systems are
solved in double precision to relative tolerance of $10^{-8}$.

\section{Conclusions and Summary}
We have carried out development effort for a few selected kernels used
in lattice QCD. The first of these efforts included the development of
a hybrid MPI/OpenMP library which includes parallelized kernels for
the Wilson Dirac operator and few associated solvers. A number of
parallelization strategies have been investigated, such as for
overlapping communication with computations. The code has been shown
to scale fairly well on the Cray XE6. In terms of single process
performance, we carried out initial vectorization efforts for AVX
where we see an improvement of $\sim$1.5 compared to the ideal 2. In
addition we have investigated several data-ordering and associated
prefetching strategies. 

For the case of tmLQCD, the main software code of the ETMC
collaboration, we have implemented an efficient linear solver which
incrementally deflated the twisted-mass Dirac operator to give a
speed-up of about 3 times when enough right-hand-sides are
required. This is already in use in production projects, such as in
Refs.~\cite{Abdel-Rehim:2013wlz} and~\cite{Alexandrou:2013nda}.

All codes are publicly available. PLQCD is available through the
\href{http://www.hpcforge.org/projects/pracelqcd/}{HPCFORGE} website
at the Swiss National Supercomputing Centre (CSCS) where more
information is available within the code documentation. Our EigCG
implementation in tmLQCD is available via \href{https://github.com/amabdelrehim/tmLQCD}{git-hub}.\\
{\bf Acknowledgements}\\
This talk was a part of a coding session sponsored partially by the
PRACE-2IP project, as part of the "Community Codes Development" Work
Package 8.  PRACE-2IP is a 7th Framework EU funded project
(http://www.prace-ri.eu/, grant agreement number: RI-283493). We would
like to thank the organizers of the 2013 Lattice meeting for their
strong support to make the coding session a success and provide all
organization support.  We would like to thank C. Urbach, A. Deuzmann,
B. Kostrzewa, Hubert Simma, S. Krieg, and L. Scorzato for very
stimulating discussions during the development of this project. We
acknowledge the computing resources from Tier-0 machines of PRACE
including JUQUEEN and Curie machines as well as the Todi machine at
CSCS. We also acknowledge the computing support from NERSC and the
Hopper machine.


\begin{thebibliography}{99} 
 \bibitem{prace-2ip-wp8} \href{http://www.prace-ri.eu/}{http://www.prace-ri.eu/}.
 \bibitem{tmLQCD-ref}K.~Jansen and C.~Urbach, \emph{Comput.Phys.Commun.} {\bf 180}, 2717 (2009), [{\tt arXiv:0905.3331}].
 \bibitem{tmLQCD-github}ETM Collaboration, \href{https://github.com/etmc/tmLQCD}{https://github.com/etmc/tmLQCD}.
 \bibitem{Chroma-software}\href{http://usqcd.jlab.org/usqcd-docs/chroma/}{http://usqcd.jlab.org/usqcd-docs/chroma/}.
 \bibitem{D8.3}See the public deliverable \href{http://www.prace-ri.eu/Public-Deliverables#PD_2IP}{D8.3} on the PRACE website under PRACE-2IP. 
 \bibitem{Albert-plenary}A.~Deuzeman, \pos{PoS(LATTICE 2013)}.
 \bibitem{Intel-developer-manual}See \href{http://www.intel.com/content/www/us/en/processors/architectures-software-developer-manuals.html}{the Intel Developer manual}.
 \bibitem{eigCG}A.~Stathopoulos and  K.~Orginos, \emph{Computing and deflating eigenvalues while solving multiple right-hand side linear
systems with an application to quantum chromodynamics}, \emph{SIAM J. Sci. Comput.} 2010; {\bf 32(1)}:439--462, [{\tt arXiv:0707.0131}].
  \bibitem{DDHMC}See for example the documentation of the \href{http://luscher.web.cern.ch/luscher/DD-HMC/}{$DD-HMC$} code by M. L$\ddot{\rm u}$scher.
 \bibitem{hopper}\href{http://www.nersc.gov/systems/hopper-cray-xe6/}{The Hopper Cray XE6 machine at NERSC}.
 \bibitem{hopper-docs-combine-omp-mpi}
 See \href{http://www.nersc.gov/users/computational-systems/hopper/performance-and-optimization/using-openmp-effectively-on-hopper/}
 {documentation for combining MPI and openMP on the NERSC website.}
 \bibitem{TMF}R.~Frezzotti {\it et al.}  [Alpha Collaboration],
  \emph{Lattice QCD with a chirally twisted mass term},
  \emph{JHEP} {\bf 0108}, 058 (2001)
  [{\tt hep-lat/0101001}].
\bibitem{Alexandrou:2013nda} 
  C.~Alexandrou, M.~Constantinou, S.~Dinter, V.~Drach, K.~Hadjiyiannakou, K.~Jansen, G.~Koutsou and A.~Vaquero,
  arXiv:1309.7768 [hep-lat].
\bibitem{Abdel-Rehim:2013wlz} 
  A.~Abdel-Rehim, C.~Alexandrou, M.~Constantinou, V.~Drach, K.~Hadjiyiannakou, K.~Jansen, G.~Koutsou and A.~Vaquero,
  arXiv:1310.6339 [hep-lat].
\end{thebibliography}
\end{document}